\documentclass[twocolumn,amsmath,amssymb,prl]{revtex4}
\usepackage{graphicx}
\usepackage{dcolumn}
\usepackage{bm}

\newcommand{\ga}{\gtrsim}
\newcommand{\sgra}{Sgr~A*~}

\def\apj{ApJ}

\def\apjl{ApJ Lett.}
\def\mnras{MNRAS}

\def\aap{A\& A}

\def\beq{\begin{equation}}
\def\eeq{\end{equation}}

\begin{document}
\input{epsf}

\title{Properties of the Radio-Emitting Gas Around SgrA*}

\author{Abraham Loeb$^{1,2}$ \& Eli Waxman$^3$}

\affiliation{$^1$ Astronomy Department, Harvard University, 60 Garden
Street, Cambridge, MA 02138, USA}

\affiliation{$^2$ Einstein Minerva center, Weizmann Institute of Science, 
Rehovot 76100, Israel,}

\affiliation{$^3$Physics Faculty, Weizmann Institute of Science, Rehovot
76\ 100, Israel}

\begin{abstract}

We show that the radial profiles of the temperature and density of the
electrons as well as the magnetic field strength around the massive black
hole at the Galactic center, \sgra, may be constrained directly from
existing radio data without any need to make prior assumptions about the
dynamics of the emitting gas.  The observed spectrum and
wavelength-dependent angular size of \sgra indicate that the synchrotron
emission originates from an optically-thick plasma of quasi-thermal
electrons. We find that the electron temperature rises above the virial
temperature within tens of Schwarzschild radii from the black hole,
suggesting that the emitting plasma may be outflowing. Constraints on the
electron density profile are derived from polarization measurements. Our
best-fit results differ from expectations based on existing theoretical
models. However, these models cannot be ruled out as of yet due to
uncertainties in the source size measurements.  Our constraints could
tighten considerably with future improvements in the size determination and
simultaneous polarization measurements at multiple wavelengths.

\end{abstract}


\maketitle

\section{I. Introduction}

The supermassive black hole at the Galactic center, \sgra, occupies the
largest angle on the sky among all known black holes.  Its extended image
provides an excellent opportunity to study the physics of low-luminosity
accretion flows.  

The bolometric luminosity of \sgra $\sim 10^{36}~{\rm erg~s^{-1}}$ is $\sim
8.5$ orders of magnitude smaller than the Eddington limit for its black
hole mass of $\sim 4\times 10^6M_\odot$.  Over the past decade various
theoretical models have been proposed to explain the low luminosity of
\sgra despite the large gas reservoir from stellar winds in its
vicinity. Among the early models invoked was an {\it Advection Dominated
Accretion Flow (ADAF)} involving hot protons and cold electrons with a low
radiative efficiency at the Bondi accretion rate of $\sim
10^{-5}M_\odot~{\rm yr^{-1}}$ \cite{Narayan}.  Subsequently, the detection
of linear polarization was used to set an upper limit on the electron
density near the black hole, which ruled out the original ADAF proposal
\cite{QG00,Ago,Mac} and favored shallower density profiles with a lower
accretion rate such as in a {\it Convection Dominated Accretion Flow
(CDAF)} \cite{QG00b}.  Later variants of the ADAF model allowed for
outflows, namely mass loss from the inflowing gas \cite{BB}.  Most
recently, an improved {\it Radiatively Inefficient Accretion Flow (RIAF)}
model was proposed, involving substantial mass loss (although the
outflowing mass is ignored in calculating the radio emission), a
non-thermal component of electrons, and different electron and proton
temperatures.  Other models associated the radio emission with a jet
\cite{Jet} or a compact torus near the black hole \cite{Liu}.

In parallel to these modelling developments, the data on \sgra has improved
dramatically over the past few years. The latest observations include new
determinations of the size, spectral luminosity, polarization and rotation
measure of the source as a function of wavelength
\cite{Shen05,Krichbaum06,Marrone07,YUAN}.  With the rich data set that is
now available, it is timely to remove any theoretical prejudice and ask:
{\it what does the data alone tell us about the properties of the radiating
gas?}  In addressing this minimal question here, we deviate from past
practice of modelers who made assumptions about the dynamics of the
accreting gas before interpreting the observational data on \sgra. We avoid
dynamical assumptions and attempt to constrain the properties of the
radio-emitting gas directly from the data itself. 

As discussed in detail in \S~II below, the measurements of the source size
at different radio wavelengths provide crucial constraints on the
properties of the gas surrounding \sgra. Current size measurements are
unfortunately subject to large error bars, which in turn imply large
uncertainties in the inferred gas properties. Our analysis provides an
estimate of the spatial dependence of gas properties based on current
measurements, adopting a frequency dependent size $r\propto\nu^\alpha$ with
$\alpha=1\pm0.3$ \citep{Shen05,Krichbaum06}. Our methodology demonstrates
how more accurate measurements may be used to obtain better constraints
with no model-dependent assumptions about the dynamics of the gas.

In the different subsections of \S~II we apply our approach to various
aspects of the data on \sgra that are currently available. We compare our
results to previous work in \S~III. Finally, \S~IV summarizes our main
conclusions.

\section{II. Empirical Constraints}

\subsection{II.1. Radio Spectrum and Size: Data}

\begin{figure}[htbp]
\includegraphics[width=8.5cm]{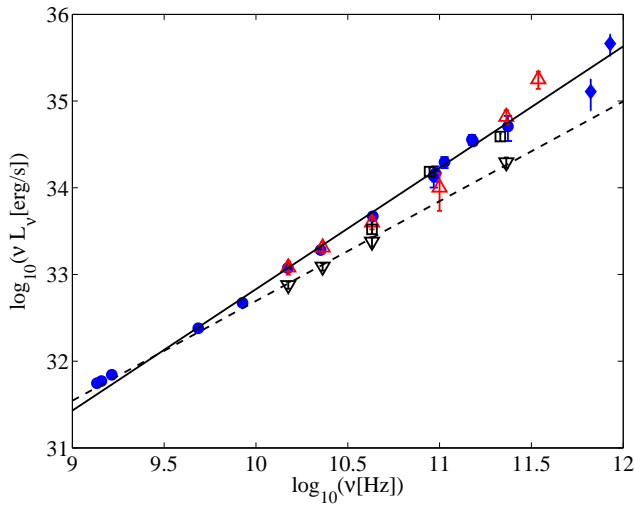}
\caption{Specific luminosity of SgrA* from simultaneous multi-frequency
data of \citet[][filled circles]{Falcke98}, and \citet[][up/down triangles
representing the flux at different times, close to the times of
maximum/minimum in the 1~mm flux]{Zhao03}. Diamonds denote flux
measurements by \citet[][666~GHz]{Zylka95} and
\citet[][850~GHz]{Serabyn97}.  Squares denote the flux densities derived
from the data used for size determination, given in
\citet{Krichbaum98,Krichbaum06} (for 1.4~mm and 3.4~mm) and in
\citet{Shen05} (for 7~mm). The solid line shows the power-law relation
$\nu L_\nu=1.7\times10^{34}\nu_{11}^{1.4}{\rm erg~s^{-1}}$, and the dashed
line is $0.7\times10^{34}\nu_{11}^{1.2}{\rm erg~s^{-1}}$. The assumed
distance to \sgra is 8kpc.}
\label{fig:flux}
\end{figure}

The spectral luminosity of SgrA* is time dependent.  As illustrated in
Fig.~\ref{fig:flux}, the observed specific luminosity $L_\nu$ per unit
frequency $\nu$ at the brighter emission epochs \cite{Falcke98,Zhao03} is
well described by a power-law form in the frequency range of 3--1000~GHz,
\begin{equation}\label{eq:L_nu}
    \nu L_\nu=1.7\times10^{34}\nu_{11}^{1.4}\,{\rm erg~s^{-1}}
\end{equation}
where $\nu_{11}\equiv (\nu/10^{11}~{\rm Hz})$.  Since the flux measured at
the times used for size determination is close to the brighter emission
values, we will use this power-law index in our phenomenological
discussion. 

We note that at low frequencies $<3$~GHz the observed flux somewhat exceeds
the flux given by Eq.~(\ref{eq:L_nu}) [equivalently, fitting a power-law
only to the data at $<10$~GHz would result in a fit that under-predicts the
flux at higher frequencies (as indicated by the dashed line in
Fig.~\ref{fig:flux}), i.e. in a "sub-mm excess"]. This deviation is of no
significance to the analysis below, which focuses on higher frequencies,
$>10$~GHz. It does imply, however, that in applying our simple analytic
results to the lowest observed frequencies, some minor quantitative
modifications would need to be introduced. As explained below, we argue
that the radio flux is dominated at different frequencies by plasma located
at different radii.  The deviation from Eq.~(\ref{eq:L_nu}) below $3$~GHz
implies therefore that the gas temperature and magnetic field strength at
large radii, $\gtrsim10^{14.5}$~cm, differ slightly from those obtained
using our simple power-law scalings, which are based on
Eq.~(\ref{eq:L_nu}).

As demonstrated in \S~II.2 below, measurements of the size of \sgra at
various radio wavelengths provide important constraints on the emitting gas
\cite{Shen05,Bower,Krichbaum06}. For a black hole mass of $M=4\times
10^6M_\odot$ \cite{Genzel}, the radial scale is set by the Schwarzschild
radius of $R_s=1.2\times 10^{12}~{\rm cm}$, which corresponds to an angle
of 0.01 mas on the sky at our adopted distance of $8$ kpc.  Table~I
presents the latest data from \citet{Krichbaum06} and compares the inferred
brightness temperature $T_b$ to the virial temperature~\footnote{For an
optically thick source with a top-hat intensity distribution, the
full-width at half-maximum (FWHM) of a Gaussian which contains the same
flux fraction as the top-hat within the FWHM size is
$\sqrt{2/\pi}\int_0^{\sqrt{2\log2}}dx\exp(-x^2/2)=0.76$ times the source
diameter. We use this result when interprating the values quoted by
\citet{Krichbaum06} for the geometric mean of FWHM of the major and minor
axes of the image of \sgra.}.  The brightness
temperature~\footnote{Throughout the paper, we set Boltzmann's constant
$k_B$ to unity and express temperatures in energy units.} at a radius $r$
from \sgra where the emissivity at an observed frequency $\nu$ peaks, is
defined through the relation
\begin{equation}
\nu L_\nu=4\pi
r^2(\nu)\times f_g\times{2\nu^3\over {c^2}}T_b(\nu), 
\label{eq:Lnu}
\end{equation}
where the geometric coefficient $f_g\le1$ is the ratio between the emission
surface area and the area of a sphere of radius $r$ ($f_g=1$ for a sphere
and $f_g=0.5$ for a two-sided disk with the same radius).   We
conservatively assume that the surface area scales as $r^2$ since this
provides the lowest brightness temperature (which, as we will show, is
already above the virial temperature at large radii).  We define the virial
temperature $T_v$ by equating the thermal kinetic energy of the plasma to
half of the gravitational potential energy per proton,
\begin{equation}
2\times \frac{3}{2}T_v={GMm_p\over 2r},
\end{equation}
where $m_p$ is the proton mass. (Note that the ``escape temperature'' at
which the thermal kinetic energy of the plasma exceeds its gravitational
potential energy is $2T_v$.)  Using Eq.~(\ref{eq:L_nu}) and
$M=4\times10^6 M_\odot$ we obtain
\begin{equation}\label{eq:Tb}
    T_b=7.6 (2f_g)^{-1}\nu_{11}^{-1.6}r_{13}^{-2}\,{\rm MeV},
\end{equation}
\begin{equation}\label{eq:Tv}
    T_v=9.5r_{13}^{-1}\,{\rm MeV},
\end{equation}
where $r_{13}\equiv (r/10^{13}~{\rm cm})$.  Table I shows that $T_b$ must
be close to $T_v$ at $r_{13}=1$, and that there is preliminary evidence 
that $T_b/T_v$ increases with radius.

\citet{Shen05} infer a power-law dependence of the intrinsic size of \sgra
on wavelength $\lambda$ of $r\propto\lambda^\beta$, with
$\beta=1.09\pm0.33$. For our phenomenological analysis we use
\begin{equation}\label{eq:size}
    r_{13}=1.0\nu_{11}^{-1/\alpha},\quad \nu_{11}=1.0 r_{13}^{-\alpha}
\end{equation}
with $\alpha\equiv \beta^{-1}\approx 1\pm0.3$, implying
\begin{equation}\label{eq:TbTv}
    2f_g{T_b\over T_v}= 0.8r_{13}^{1.6\alpha-1}.
\end{equation}

\begin{table}[hdtp]
\caption{Measured size and brightness temperature of \sgra. The listed
radii are related to the FWHM intrinsic sizes of \citet{Krichbaum06} by
$r=0.5\times{\rm FWHM}/0.76$ (see footnote [36]).} 

\begin{center}
\begin{tabular}{||c|c|c|c|c|c||}
\hline\hline
{$\lambda$ [mm]} &
{$\nu_{11}$} & {$r_{13}$} & {$2f_gT_b$ [MeV]} &
{$T_v$ [MeV]} & {$2f_gT_b/T_v$} \\ \hline
 & & & & & \\
 1.4 & 2.1 &
$0.86\pm0.47$ & $2.4^{+9}_{-1.4}$ & $11$ 
& $0.2^{+0.2}_{-0.08}$ \\ 3.4 & 0.88 & $1.2\pm0.23$ &
$7.2^{+4}_{-2.2}$ & $8.0$ & $0.8^{+0.2}_{-0.14}$ \\ 7 & 0.43 &
$2.0\pm0.55$ & $4.8^{+4.4}_{-1.8}$ & $4.8$ & $1.0^{+0.4}_{-0.2}$ \\
\hline \hline
\end{tabular}
\end{center}
\label{table1}
\end{table}

\subsection{II.2. Radio Spectrum and Size: Implications}

The observed radio luminosity originates most likely from synchrotron
emission by relativistic electrons \cite{Goldston}. The frequency
dependence of source size implies that the emission cannot originate from
an optically-thin plasma. To see this, consider the electrons at $r$
dominating the emission at a frequency $\nu(r)$. If the optical depth is
small, then these electrons would produce a spectrum $\nu L_\nu\propto
\nu^{4/3}$ at $\nu<\nu(r)$, close to the observed spectrum at these
frequencies. This implies that the emission at $\nu<\nu(r)$ would be
dominated by electrons at $r$, which is inconsistent with the frequency
dependence of source size. We therefore conclude that the optical depth for
synchrotron self-absorption satisfies $\tau_\nu[\nu(r),r]\ge1$. For a
similar reason, the characteristic synchrotron emission frequency
$\nu_c(r)=\langle \gamma_e^2\rangle (eB/2\pi m_ec)$ of the electrons at $r$
[dominating the emission at a frequency $\nu(r)$] must satisfy
$\nu_c(r)\approx\nu(r)$.  If $\nu_c(r)\gg\nu(r)$ then for
$\tau_\nu[\nu(r),r]\gg1$ these electrons would produce a flux $\nu
L_\nu\propto \nu^{3}$ at $\nu>\nu(r)$, inconsistent with the observed
spectrum, and for $\tau_\nu[\nu(r),r]=1$ these electrons would produce a
flux $\nu L_\nu\propto \nu^{4/3}$ at $\nu>\nu(r)$, dominating the emission
at $\nu>\nu(r)$ in conflict with the frequency dependence of source size.
    
We therefore conclude that radiation at different radii $r$ is dominated by
electrons with $\nu_c(r)\approx\nu(r)$ and that $\tau_\nu[\nu(r),r]\ge1$.
Next, we argue that the electron energy distribution may be characterized
by a single energy or an effective temperature $T_e$. That is, we show that
the energy distribution of electrons cannot be highly
non-thermal. Consider, for example, a power-law distribution of electron
energies, $dn_e/d\gamma_e\propto\gamma_e^{-p}$. Such a distribution would
be consistent with observations provided that $\tau_\nu[\nu(r),r]\approx1$
(rather than $\tau_\nu[\nu(r),r]\gg1$), since otherwise the flux emitted by
electrons at $r$ would extend beyond $\nu(r)$ as $\nu
L_\nu\propto\nu^{7/2}$, exceeding the observed flux. For
$\tau_\nu[\nu(r),r]\approx1$, $\nu L_\nu\propto\nu^{(3-p)/2}$ at
$\nu>\nu(r)$, and the value of $p$ is constrained by the ratio between the
far-infrared luminosity, $\simeq3\times 10^{34}$~erg/s at a frequency $\sim
10^{14}$~Hz \cite{Genzel03,Ghez04}, and the radio luminosity,
$\simeq5\times10^{35}$~erg/s at $\sim 10^{12}$~Hz, to be $p\ge4.3$.  This
large power-law index implies that only a small fraction of the total
energy can be carried by electrons of energy exceeding that of the
electrons dominating the radio emission. Moreover, we will show in \S~II.3
that the extension of such a power-law to electron energies much below that
of the electrons dominating the radio emission would imply a very large
rotation measure, inconsistent with observations~\footnote{One may
postulate, of course, a power-law distribution with $p<4.3$, which cuts off
just above the energy of the electrons we observe (so as to avoid over
producing the $10^{14}$~Hz flux). However, such a cut-off would be
physically unnatural.}. These constraints are satisfied by recent RIAF
models \cite{YQN04,YUAN} which associate only a small fraction of the total
electron energy with a power-law component. However, our simple analysis
shows that if thermal emission at different frequencies originates at
different radii to account for the observed spectrum and size measurements,
then there is no need for an ad-hoc non-thermal component.

Since the emission originates from an optically thick plasma, the
characteristic temperature (energy) $T_e$ of the electrons dominating the
radiation is
\begin{equation}\label{eq:TeTb}
T_e(r)\approx T_b(r)\approx7.6 (2f_g)^{-1}r_{13}^{1.6\alpha-2}\,{\rm MeV}.     
\end{equation}
The electron temperature has to satisfy 
\begin{equation}
\nu_c(r)=12\left({T_e(r)\over m_e c^2}\right)^2\times
0.3{eB(r)\over 2\pi m_e c}=\nu(r),
\end{equation}
where $m_e$ is the electron mass and we used the relation
$\langle\gamma_e^2\rangle=12[T_e(r)/m_e c^2]^2$ in which angular brackets
denote an average over a relativistic Maxwellian of temperature
$T_e$. Based on Eqs.~(\ref{eq:size}) and (\ref{eq:Tv}) this requirement
implies
\begin{equation}\label{eq:nu_c}
    \left[\frac{T_e(r)}{T_v(r)}\right]^2 B = 27 r_{13}^{2-\alpha}\,{\rm G}.
\end{equation}
In deriving this result we have not used the inferred source size but only
the fact that it is frequency dependent. Using the size estimate in
Eq.~(\ref{eq:TbTv}) and $T_e\approx T_b$, we then get
\begin{equation}\label{eq:B}
    B=27(2f_g)^{2}r_{13}^{4-4.2\alpha}\,{\rm G}.
\end{equation} 

The $\nu(r)\propto r^{-\alpha}$ scaling of the observed radiation frequency
on emission radius implies $T_e\approx
T_b=5\nu_{11}^{-1.6+2/\alpha}$~MeV. In order for the synchrotron model to
hold down to $\sim1$~GHz, the electrons must remain relativistic, i.e. the
condition $5\times10^{-2(-1.6+2/\alpha)}>1$ must hold, implying
$1/\alpha\le1$ or $\alpha\ge 1$. It therefore appears that the value
$1/\alpha\approx1$ of \citet{Shen05} is preferred over the alternative
suggestion for higher values $1/\alpha\approx1.5\pm0.2$ \cite{Bower}.

The results in Eqs.~(\ref{eq:TbTv}), (\ref{eq:TeTb}) and (\ref{eq:B}) have
several important implications. First, $T_e$ is close to $T_v$ at
$r_{13}\sim1$, and $T_e/T_v$ increases with radius approximately as
$r^{1/2}$ for $\beta\sim 1.1$ \footnote{Note that for $\alpha=1/1.6=0.625$,
which might still be consistent with observations \cite{Bower},
$T_e/T_v=const$, but then $B^2\propto r^{2.75}$ which is physically
implausible.}. This implies that the gas cannot be confined to a thin disk
and the flow geometry must be quasi-spherical.  Moreover the radio-emitting
gas is not likely to be flowing in but rather flowing out since its thermal
kinetic energy exceeds the gravitational binding energy beyond a radius of
a few tens of Schwarzschild radii ($r_{13}\ga 2$). Our conclusions
would only be strengthened if the emitting plasma follows
a jet geometry for which the surface area scales as $r^\delta$
with $\delta<2$ [see the discussion following Eq. (\ref{eq:Lnu})].


Finally, we note that the plasma under consideration is collisionless as
the Coulomb collision time is much longer than the dynamical time of the
gas $2\pi{\sqrt{r^3/GM}}=0.9\times 10^4 r_{13}^{3/2}$s. However, collective
plasma effects should operate \cite{Sharma} since the inverse of the plasma
frequency or electron gyro-frequency are much shorter than the dynamical
time.  We use the term ``temperature'' in our discussion to characterize
the typical electron energy even if the electron distribution function
happens to be non-Maxwellian.

\subsection{II.3. Density constraints: Opacity, Rotation Measure and Circular Polarization}

The optical depth to synchrotron self-absorption is
\begin{equation}\label{eq:tau}
    \tau[\nu(r),r]=\alpha_\nu r j_\nu=\frac{c^2}{2\nu^2
    T_e(r)}r\frac{n_e e^3 B}{m_e c^2} \propto n_e r^{7-3.8\alpha},
\end{equation}
giving
\begin{equation}\label{eq:nB}
    n_e B= 2.0\times10^5 (2f_g)^{-1}\tau(r) r_{13}^{-3-0.4\alpha}\,{\rm
    G~cm^{-3}},
\end{equation}
and
\begin{equation}\label{eq:n}
    n_e= 5.6\times10^3 (2f_g)^{-3}\tau(r)r_{13}^{3.8\alpha-7}\,{\rm
    cm}^{-3},
\end{equation}
For $\alpha\approx 1$ the optical depth increases with radius,
$\tau\propto n_e r^{3}$. In order to ensure that $\tau(r)>1$, it is
sufficient to require that this condition will hold at $r_{13}=1$, implying
$n_e(r_{13}=1)>10^3{\rm cm}^{-3}$.

The relativistic rotation measure of a fluid of electrons with a thermal
Lorentz factor $\gamma_e$ and density $n_e$ threaded by a coherent magnetic
field ${\bf B}$, is given by $RRM=8\times10^5(n_e/{\rm cm}^{-3})(B/{\rm
G})(r/{\rm pc})\langle \gamma_e^{-2}\rangle \,{\rm rad~m^{-2}}$
\cite{QG00}.  As long as ${\bf B}$ is coherent, we may express this
rotation measure in terms of $\tau$ as
\begin{eqnarray}
    RRM&\approx& 2.5\times10^3 (2f_g)\tau(r)r_{13}^{2-3.6\alpha}\,{\rm
    rad~m^{-2}} \label{eq:RRM1}
    \\ &=&5\times10^5 (2f_g)^4 n_{e,6}
    r_{13}^{9-7.4\alpha}\,{\rm rad~m^{-2}},
\label{eq:RRM}
\end{eqnarray}
where we substituted $\langle \gamma^{-2}\rangle \approx (m_ec^2/T)^2$.
For $\alpha\approx1.1$, the rotation measure scales as $RRM\propto n_e r$
and may be either decreasing or increasing with $r$. In the latter case,
the rotation measure is dominated at all frequencies by the same outermost
electron shell, while in the former case it is dominated by electrons near
the radius where radiation is emitted (and so it is expected to be larger
for higher frequencies or smaller radii). The observed rotation measure of
$\sim 6\times10^5\,{\rm rad/m^2}$ at a frequency of $\sim2\times10^{11}$~Hz
\citep[Ref.][and references therein]{Marrone07}, implies for a coherent
${\bf B}$-field that
\begin{equation}\label{eq:n_max}
    n_e\le10^6 (2f_g)^{-4}r_{13}^{7.4\alpha-9}{\rm cm}^{-3}.
\end{equation}
\citet{Marrone07} report measurements at $2.3\times10^{11}$~Hz and
$3.5\times10^{11}$~Hz. While there is an indication that the rotation
measure is higher at the higher frequency by a factor of few (see their
Table I) the observations at the two frequencies are not simultaneous, and
since the source is variable the differences may be due to variability. If
the rotation measure differences are real and not due to the temporal
variability, Eq.~(\ref{eq:RRM}) requires
\begin{equation}\label{eq:n13}
    n_e(r_{13}=1)\sim10^{6}(2f_g)^{-4}{\rm cm}^{-3} ,
\end{equation}
with the density decreasing with $r$ at least as steeply as
$r^{-9+7.4\alpha}$. For this density (and $f_g\sim1/2$), the magnetic field
and thermal energy densities are comparable at $r_{13}=1$ and
$\tau(r_{13}=1)\sim10^2$.  A turbulent magnetic field would generate a
random walk in the net rotation measure and so the electron density
inferred from the $RRM$ observations would increase by the square-root of
the number of field reversals (coherent ${\bf B}$ patches) along the region
where the $RRM$ originates. The large linear polarization observed at
frequencies $\nu >10^2$GHz for \sgra implies that the inferred magnetic
field is not highly tangled in the innermost region. The low level of
linear polarization at lower frequencies is consistent with the notion that
the emission at different frequencies originates from different radii.

For $\alpha\approx1$ the electrons become mildly relativistic at
$r\sim10^{15}$~cm, where the emission is predicted to peak around
$\sim1$~GHz. This may account for the circular polarization observed at
these low frequencies \cite{Bower99,Sault99,BowerC}.

Finally, we note that (as mentioned in \S~II.2) the observed $RRM$ excludes
a power-law extension of the electron energy distribution,
$dn_e/d\gamma_e\propto\gamma_e^{-p}$ with $p\ge4.3$, down to energies
significantly lower than $T_e$. The contribution of lower energy electrons
to the rotation measure is proportional to $n_e/\gamma_e^2\propto \gamma_e
(dn_e/d\gamma_e)/\gamma_e^2 \propto \gamma_e^{-p-1}$. Using
Eq.~(\ref{eq:RRM1}) and denoting by $\gamma_m$ the minimum electron Lorentz
factor we have $RRM\approx2.5\times10^3\tau(T_e/\gamma_m m_e c^2)^{-p-1}
\,{\rm rad/m^2}$, which implies for $p>4.3$ that the rotation measure would
exceed the observed value of $\simeq5\times10^5\,{\rm rad/m^2}$ for
$\gamma_m m_e c^2/T_e\le1/2$.

\subsection{II.4. Equipartition and Entropy}

The equipartition ratio between the magnetic energy density and the thermal
energy density of the electrons scales as
\begin{equation}\label{eq:ep}
    \frac{B^2/8\pi}{{3\over 2}n_eT_e}\propto
    \frac{r^{10(1-\alpha)}}{n_e},
\end{equation}
and the entropy scales as
\begin{equation}\label{eq:s}
    \frac{T_e^3}{n_e}\propto\frac{r^{4.8\alpha-6}}{n_e}.
\end{equation}
Requiring uniform entropy and equipartition fraction gives
$\alpha=16/14.8=1.08$ which is surprisingly within the range inferred by
\citet{Shen05}. This special value yields the scalings $T_e\propto
r^{-0.27}$, $n_e\propto r^{-0.82}$, and $B\propto r^{-0.54}$.  Substituting
these power-law scalings in Eq. (\ref{eq:RRM}) implies that the rotation
measure $RRM$ is nearly independent of emission radius or observed
frequency. This result can be tested by future observations that would
monitor the time dependence of $RRM$ at different frequencies \cite{Mar}.
A similar time dependence at different frequencies would imply that the
rotation measure is dominated by a common outer shell.

\section{III. Comparison with earlier work}

The RIAF models generically predict that the radio emission is dominated by
thermal electrons near the innermost stable circular orbit (ISCO) in the
central region of the accretion disk \citep{YQN03}. The radio spectrum
produced in this model by the thermal electrons is inconsistent with the
observed spectrum, which is well described by the power-law form in
Eq.~(\ref{eq:L_nu}). A power-law electron component is therefore added to
the thermal RIAF component in an ad-hoc manner, where the spectral index
and normalization of the power-law component are tailored to match the low
frequency radio data \cite{YQN04,YUAN}.  We have pointed out in \S~II.2
that the size measurements indicate that the source size is frequency
dependent, implying that the radiation is dominated at different
frequencies by electrons at different locations. This in turn suggests that
the observed spectrum reflects the spatial dependence of electron
temperature rather than the energy distribution of the electrons at a
single radius.

Since the emission of radiation in RIAF models is dominated by the
innermost region of the disk, at $r\sim 3R_s=3.6\times10^{12}$~cm, the
observed size is dominated in these models by foreground interstellar
scattering at all frequencies. The addition of a power-law electron
component to these models increases slightly the intrinsic source size
\cite{Yuan06}, but does not change the requirement that the measured source
size will be dominated at all frequencies by interstellar medium
scattering. These results appear to be at odds with the latest size
measurements which indicate that the intrinsic size is resolved well beyond
the expected level of interstellar image broadening at $\lambda\le3.5$~mm
\cite{Krichbaum06} and that the intrinsic size is smaller at higher
frequencies.

As mentioned in the Introduction, existing size measurements are subject to
large uncertainties. While our best-fit profiles disfavor existing models,
such models cannot be ruled out based on current data. Future, more
accurate, measurements will allow us to draw more decisive conclusions.
Our current analysis underlines the importance of future improvements in
the size measurements, and provides a methodology for interpreting future
results.

\section{IV. Summary}

We have shown in \S~II.2 that measurements of the spectrum and of the
wavelength-dependent size of the radio emission from \sgra indicate that
this emission is dominated by optically-thick quasi-thermal plasma, and
that the observed spectrum reflects the spatial dependence of the electron
temperature (rather than the energy distribution of electrons at some
particular radius). We have derived the electron temperature and magnetic
field profiles [Eqs.~(\ref{eq:TeTb}) and~(\ref{eq:B})], and found that the
electron temperature increases above the virial temperature beyond a
distance of a few tens of Schwarzschild radii from the black hole.  The
observed rotation measure was then used to constrain the density profile
[Eqs.~(\ref{eq:n_max}) and~(\ref{eq:n13})].  The low density inferred for
the gas near \sgra could in principle be accounted for by winds from the
innermost S-stars \cite{Loeb}.  Although we have not proposed a dynamical
model for the plasma, we have pointed out in \S~II.4 that observations are
consistent with an isentropic gas profile and equipartition magnetic field.

Our results imply that the radio emitting gas cannot be confined to a thin
disk and the flow geometry must be geometrically thick. Moreover, the
radio-emitting gas is not likely to be inflowing but rather outflowing,
since its thermal kinetic energy exceeds the gravitational binding energy
beyond a radius of a few tens of Schwarzschild radii [$r_{13}\ga 2$, see
Table I and Eq.~(\ref{eq:TbTv})]. There may also be a colder accreting
component that is sub-dominant in terms of its synchrotron emission.  Such
a component would likely be confined to a thinner disk geometry that would
have only a limited effect on the rotation measure of the radiation emitted
by the hot outflowing atmosphere above it. (However, if the cold component
is optically-thick then it would make the image of SgrA* asymmetric at a
level that would depend on the inclination of the disk. The resulting
frequency dependence of the image centroid location could be constrained by
observations.) An electron temperature profile which does not decline with
increasing radius as fast as the virial temperature does, would be
consistent with an adiabatic outflow in which the electrons are hotter than
the protons because their relativistic temperature declines with decreasing
density as $T \propto n^{\Gamma-1}$ with an adiabatic index ($\Gamma=4/3$)
that is smaller than that of the protons (5/3). Heat conduction could also
help to flatten the electron temperature profile \cite{Menou}.

The large uncertainty in the inferred source size, reflected by the large
uncertainty in the index $\alpha=1\pm0.3$ of the relation
$r\propto\nu^\alpha$, translates to a large uncertainty in the temperature
and magnetic field profiles, as implied by Eqs.~(\ref{eq:B})
and~(\ref{eq:TeTb}).  Our preliminary conclusions from existing data differ
from current theoretical models \cite{YUAN}, although not at a
statistically robust level. These potential discrepancies provide added
incentive to obtain better data through future observations. 

Finally, we note that the uncertainty in the determination of the density
profiles is related not only to uncertainties in the source size
measurements, but also to the lack of simultaneous multi-frequency
measurements of the rotation measure (\S~II.3). An accurate determination
of the density profile would require therefore not only accurate size
measurements, but also simultaneous multi-frequency measurements of the
rotation measure. Ultimately, direct imaging of \sgra with a Very Large
Baseline Array at sub-mm wavelengths \cite{Falcke,Broderick,Shen-VLBI}
would resolve the accretion flow near the black hole ISCO and unravel
unambiguously the properties of the emitting gas there.

\bigskip
\bigskip
\bigskip
\bigskip

\paragraph*{Acknowledgments}

We thank Avery Broderick, Dan Marrone, and an anonymous referee for useful
comments on the manuscript. A.L. thanks the Weizmann Institute for its kind
hospitality when this work was conducted. This work was supported in part
by ISF and Minerva grants (E. W.) and the BSF foundation (A. L.).


\begin{references}

\bibitem[Agol(2000)]{Ago} Agol, E.\ 2000, \apjl, 538, L121 


\bibitem[Blandford \& Begelman(1999)]{BB} Blandford, R.~D., \& Begelman,
M.~C.\ 1999, \mnras, 303, L1

\bibitem[Bower et al.(1999)]{Bower99} Bower, G.~C., Falcke, H., 
\& Backer, D.~C.\ 1999, \apjl, 523, L29 

\bibitem[Bower(2003)]{BowerC} Bower, G.~C.\ 2003, Astrophys. \& Space Sci.,
288, 69

\bibitem[Bower(2006)]{Bower} Bower, G.~C.\ 2006, Journal of Physics
Conference Series, 54, 370; Bower, G.~C., Goss, W.~M., Falcke, H., Backer,
D.~C., \& Lithwick, Y.\ 2006, \apjl, 648, L127

\bibitem[Broderick \& Loeb(2006)]{Broderick} Broderick, A.~E., \& Loeb, A.\
2006, Journal of Physics Conference Series, 54, 448 [astro-ph/0607279];
2006, \mnras, 367, 905 

\bibitem[Eckart et al.(2004)]{Genzel} Eckart, A., Genzel, R., 
\& Sch{\"o}del, R.\ 2004, Progress of Theoretical Physics Supplement, 155, 
159 

\bibitem[Falcke et al.(1998)]{Falcke98} Falcke, H., Goss, W.~M., Matsuo,
H., Teuben, P., Zhao, J.-H., \& Zylka, R.\ 1998, \apj, 499, 731

\bibitem[Falcke et al.(2000)]{Falcke} Falcke, H., Melia, F., \& Agol, E.\
2000, \apjl, 528, L13

\bibitem[Genzel et al.(2003)]{Genzel03} 
  Genzel, R., Sch{\"o}del, R., Ott, T., Eckart, A., Alexander, T., Lacombe, 
  F., Rouan, D., \& Aschenbach, B.\ 2003, \nat, 425, 934 

\bibitem[Ghez et al.(2004)]{Ghez04} 
  Ghez, A.~M., et al.\ 2004, \apjl, 601, L159 

\bibitem[Goldston et al.(2005)]{Goldston} Goldston, J.~E., 
Quataert, E., \& Igumenshchev, I.~V.\ 2005, \apj, 621, 785 

\bibitem[Krichbaum et al.(1998)]{Krichbaum98}
  Krichbaum, T.~P., et al.\ 1998, \aap, 335, L106

\bibitem[Krichbaum et al.(2006)]{Krichbaum06} 
  Krichbaum, T.~P., Graham, D.~A., Bremer, M., Alef, W., Witzel, A., Zensus, J.~A., \& Eckart, A.\ 2006, 
  Journal of Physics Conference Series, 54, 328 [astro-ph/0607072]

\bibitem[Liu \& Melia(2003)]{Liu} Liu, S., \& Melia, F.\ 2003,
Astronomische Nachrichten Supplement, 324, 475

\bibitem[Loeb(2004)]{Loeb} Loeb, A.\ 2004, \mnras, 350, 725

\bibitem[Macquart et al.(2006)]{Mac} Macquart, J.-P., Bower, G.~C., Wright,
M.~C.~H., Backer, D.~C., \& Falcke, H.\ 2006, \apjl, 646, L111

\bibitem[Marrone et al.(2007)]{Marrone07} 
  Marrone, D.~P., Moran, J.~M., Zhao, J.-H., \& Rao, R.\ 2007, \apjl, 654, L57 

\bibitem[Marrone(2007)]{Mar} Marrone, D.~P.\ 2007, private communication

\bibitem[Tanaka \& Menou(2006)]{Menou} Gruzinov, A. 2008, preprint
astro-ph/9809265; Tanaka, T., \& Menou, K.\ 2006, \apj, 649, 345

\bibitem[Narayan et al.(1998)]{Narayan} Narayan, R., Mahadevan, R., \&
Quataert, E.\ 1998, Theory of Black Hole Accretion Disks, 148; Quataert,
E., Narayan, R., \& Reid, M.~J.\ 1999, \apjl, 517, L101

\bibitem[Quataert \& Gruzinov(2000a)]{QG00} Quataert, E., \& Gruzinov, A.\
2000, \apj, 545, 842

\bibitem[Quataert \& Gruzinov(2000b)]{QG00b} Quataert, E., \& Gruzinov, A.\
2000, \apj, 539, 809; Narayan, R., Igumenshchev, I.~V., \& Abramowicz,
M.~A.\ 2000, \apj, 539, 798

\bibitem[Sault \& Macquart(1999)]{Sault99} Sault, R.~J., \& 
Macquart, J.-P.\ 1999, \apjl, 526, L85 


\bibitem[Serabyn et al.(1997)]{Serabyn97} 
  Serabyn, E., Carlstrom, J., Lay, O., Lis, D.~C., Hunter, T.~R., \& Lacy, J.~H.\ 1997, \apjl, 490, L77 

\bibitem[Sharma et al.(2006)]{Sharma} Sharma, P., Hammett, G.~W., Quataert,
E., \& Stone, J.~M.\ 2006, \apj, 637, 952

\bibitem[Shen et al.(2005)]{Shen05} 
  Shen, Z.-Q., Lo, K.~Y., Liang, M.-C., Ho, P.~T.~P., \& Zhao, J.-H.\ 2005, \nat, 438, 62 

\bibitem[Shen(2005)]{Shen-VLBI} Shen, Z.-Q.\ 2005, Journal of 
Korean Astronomical Society, 38, 261 

\bibitem[Yuan(2006)]{YUAN} Yuan, F. 2006, Journal of Physics Conference
Series, 54, 427 [astro-ph/0607123]

\bibitem[Yuan et al.(2002)]{Jet} Yuan, F., Markoff, S., \& Falcke, H.\
2002, \aap, 383, 854; Falcke, H.\ \& Markoff, S. 2000, Astron. \&
Astrophys., 362, 113

\bibitem[Yuan et al.(2003)]{YQN03} 
  Yuan, F., Quataert, E., \& Narayan, R.\ 2003, \apj, 598, 301 

\bibitem[Yuan et al.(2004)]{YQN04} 
  Yuan, F., Quataert, E., \& Narayan, R.\ 2004, \apj, 606, 894 

\bibitem[Yuan et al.(2006)]{Yuan06} 
  Yuan, F., Shen, Z.-Q., \& Huang, L.\ 2006, \apjl, 642, L45  

\bibitem[Zhao et al.(2003)]{Zhao03} Zhao, J.-H., Young, K.~H., Herrnstein,
R.~M., Ho, P.~T.~P., Tsutsumi, T., Lo, K.~Y., Goss, W.~M., \& Bower, G.~C.\
2003, \apjl, 586, L29

\bibitem[Zylka et al.(1995)]{Zylka95} 
  Zylka, R., Mezger, P.~G., Ward-Thompson, D., Duschl, W.~J., \& Lesch, H.\ 1995, \aap, 297, 83 





\end{references}
\end{document}